\documentclass[preprint,11pt,times]{elsarticle}

\usepackage{amsmath,amsfonts}
\usepackage{algorithmic}
\usepackage{array}

\usepackage{textcomp}
\usepackage{stfloats}
\usepackage{url}
\usepackage{verbatim}
\usepackage{graphicx}
\usepackage{subcaption}
\usepackage{soulutf8}
\usepackage[dvipsnames,svgnames,table]{xcolor}
\usepackage[colorlinks=true,allcolors=blue,final,urlcolor=black]{hyperref}
\usepackage{booktabs}
\usepackage{orcidlink}

\usepackage{balance}

\usepackage{bm}
\usepackage{mathtools}
\usepackage{nicefrac}
\usepackage{xspace}
\usepackage{enumitem}
\usepackage[many]{tcolorbox}
\usepackage{longtable}
\usepackage{wrapfig}

\definecolor{lightgray}{gray}{0.9}

\tcbset{
    lavenderbox/.style={
        colback=Lavender!30,
        colframe=Lavender!80!black!50,
        coltitle=black,
        fonttitle=\itshape,
        boxrule=0.5pt,
        arc=2pt,
        left=3pt,
        right=3pt,
        top=-2pt,
        bottom=0pt,
        breakable,
        enhanced
    },
    beigebox/.style={
        colback=black!2!white,
        colframe=black!20,
        coltitle=black,
        fonttitle=\itshape,
        boxrule=0.5pt,
        arc=2pt,
        left=3pt,
        right=3pt,
        top=-2pt,
        bottom=0pt,
        breakable,
        enhanced
    },
    stonebox/.style={
        colback=blue!5!gray!20,
        colframe=blue!30!black,
        coltitle=black,
        fonttitle=\itshape,
        boxrule=0.5pt,
        arc=2pt,
        left=3pt,
        right=3pt,
        top=-2pt,
        bottom=0pt,
        breakable,
        enhanced
    },
    mutedbluebox/.style={
        colback=cyan!5!white,
        colframe=cyan!40!black,
        coltitle=black,
        fonttitle=\itshape,
        boxrule=0.5pt,
        arc=2pt,
        left=3pt,
        right=3pt,
        top=-2pt,
        bottom=0pt,
        breakable,
        enhanced
    },
    linedbox/.style={
        colback=white,
        colframe=black!20,
        coltitle=black,
        fonttitle=\itshape,
        boxrule=0.2pt,
        arc=0pt,
        left=3pt,
        right=3pt,
        top=-2pt,
        bottom=0pt,
        breakable,
        enhanced
    },
    nobox/.style={
        colback=white,
        colframe=black!20,
        coltitle=black,
        fonttitle=\itshape,
        boxrule=0.3pt,
        arc=2pt,
        left=3pt,
        right=3pt,
        top=-2pt,
        bottom=0pt,
        breakable,
        enhanced
    },
    noboxblank/.style={
        colback=white,
        colframe=white,
        coltitle=black,
        fonttitle=\itshape,
        boxrule=0.3pt,
        arc=2pt,
        left=3pt,
        right=3pt,
        top=-2pt,
        bottom=0pt,
        breakable,
        enhanced
    },
}

\newboolean{authnotes}
\setboolean{authnotes}{true} 

\ifthenelse{\boolean{authnotes}}
{
    \DeclareRobustCommand{\tamme}[1]{{\sethlcolor{blue!30}\hl{Tamme: #1}}}
    \DeclareRobustCommand{\paul}[1]{{\sethlcolor{green!30}\hl{Paul: #1}}}
    \DeclareRobustCommand{\td}[1]{{\sethlcolor{red!30}\hl{\normalfont TODO: #1}}}
    \newcommand{\needref}[1]{\textcolor{red}{[REF #1]}}
}
{
    \DeclareRobustCommand{\tamme}[1]{}
    \DeclareRobustCommand{\paul}[1]{}
    \DeclareRobustCommand{\td}[1]{}
    \newcommand{\needref}[1]{}
}

\newcommand{\R}{\mathbb{R}}

\newcommand{\ovar}{\bm{\lambda}}
\newcommand{\feasset}{\Lambda}

\newcommand{\feasibleset}{\feasset}
\newcommand{\optvar}{\ovar}

\newcommand{\ie}{i\/.\/e\/.,\/~}
\newcommand{\eg}{e\/.\/g\/.,\/~}
\newcommand{\cf}{cf\/.\/~}

\newcommand{\Routes}{\mathcal{R}}
\newcommand{\Occs}{\mathcal{O}}

\DeclareMathOperator*{\argmax}{arg\,max}
\DeclareMathOperator*{\argmin}{arg\,min}

\newcommand{\OPT}{\hyperref[OPT]{$\textbf{PTRAP}$}\xspace}

\newcommand{\eqrefOPT}[1]{(\hyperref[#1]{\textbf{PTRAP}-\ref*{#1}})}

\begin{document}

\begin{frontmatter}

\setlength{\abovedisplayskip}{6pt}
\setlength{\belowdisplayskip}{6pt}
\setlength{\textfloatsep}{10pt plus 2pt minus 2pt}
\setlength{\intextsep}{10pt plus 2pt minus 2pt}
\setlength{\parskip}{0.4em} 
\addtolength{\topmargin}{-0.5cm} 

\title{Utilizing Bayesian Optimization for Timetable-Independent Railway Junction Performance Determination}

\author[via]{Tamme Emunds\,\orcidlink{0000-0002-4862-1872}\corref{cor1}}
\ead{emunds@via.rwth-aachen.de}

\author[dsme]{Paul Brunzema\,\orcidlink{0000-0003-3514-7339}}

\author[dsme]{Sebastian Trimpe\,\orcidlink{0000-0002-2785-2487}}

\author[via]{Nils Nießen\,\orcidlink{0000-0001-6236-8335}}

\address[via]{Institute of Transport Science, RWTH Aachen University, Aachen, Germany}
\address[dsme]{Institute for Data Science in Mechanical Engineering, RWTH Aachen University, Aachen, Germany}

\cortext[cor1]{Corresponding author}

\begin{abstract}
The efficiency of railway infrastructure is significantly influenced by the mix of trains that utilize it, as different service types have competing operational requirements.
While freight services might require extended service times, passenger services demand more predictable schedules.
Traditional methods for addressing long-term traffic assignment problems often rely on fixed-value capacity limitations, determined based on specific assumptions about traffic composition. 
This paper introduces a methodology for determining timetable-independent capacity within the traffic rate assignment problem, enabling the calculation of junction capacities under dynamic traffic distributions.
We solve the underlying non-linear constrained optimization problem maximizing the traffic throughput using Bayesian optimization (BO).
This setting combines a known objective function with expen\-sive-to-compute capacity constraints, motivating an adaption of standard BO problems, where objective functions are usually unknown.
We tailor the acquisition process in BO to this specific setting and increase performance by incorporating prior knowledge about the shape of the constraint functions into the Gaussian process surrogate model.
Our derived approaches are benchmarked on a railway junction near Paris, significantly outperforming fixed traffic composition models and highlighting the benefits of dynamic capacity allocation.
\end{abstract}


\end{frontmatter}

\section{Introduction}\label{sec:intro}

\begin{figure*}[!t]
    \centering
    \includegraphics[width=1.\textwidth]{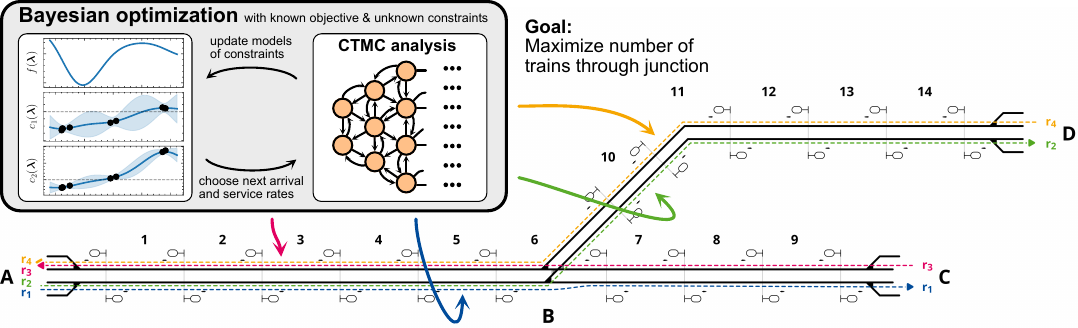}
    \caption{
    \textit{Overview of the proposed method.}
    We aim to maximize the number of trains through a junction by using Bayesian optimization to find globally optimal traffic rate assignments respecting computationally expensive to evaluate capacity constraints that are 
    modeled with
    continuous-time Markov chains. 
    }
    \label{fig:example_junction}
\end{figure*}

The global need for effective transportation modes is accelerating the development and enhancement of railway transportation networks.
To effectively allocate available resources, infrastructure managers must assess the performance of all elements within their railway networks.
Railway infrastructure is designed to last multiple decades, and infrastructure dimensioning takes place in very early planning stages.
Therefore, it is crucial to assess infrastructure performance independent of timetables, which may be fixed only a couple of years in advance.
However, the performance capabilities of railway infrastructure depend heavily on traffic distribution patterns, including the allocated rolling stock units and microscopic routes.
While recent research \cite{emundsEvaluatingRailwayJunction2024} introduced a timetable-independent method for evaluating the performance of railway junction infrastructure, its traffic distribution remained fixed for every considered number of trains.
Long-term planning problems---such as frequency assignment to traffic types and routes, or optimal line-plan determination---require methods that can accommodate dynamic traffic distributions.

This work introduces an approach for incorporating non-linear timetable capacity constraints into traffic rate assignment problems.
With this, we can describe the influence of dynamic traffic rate assignments onto the performance capability of railway infrastructure.
We solve the resulting non-linear global optimization problem---characterized by a known objective function with computationally expensive capacity constraints---using Bayesian optimization (BO).
We adapt constrained BO algorithms, which assume black-box (\ie unknown) objective functions and constraints, to our setting where the objective function is known but the capacity constraints require expensive computation and are therefore treated as black-box functions.
Furthermore, we embed domain knowledge about the constraint function's shape into the Gaussian process (GP) surrogate models that approximate the capacity constraints, resulting in improved performance.
In summary, our main contributions are:
\begin{enumerate}[label=\textit{(\roman*)}]
    \item A novel method for efficiently determining the timetable-independent capacity of railway junctions under dynamic traffic distributions.
    \item Adapting classic constrained BO algorithms to the setting of an known objective function and prior knowledge of an exponential trend of black-box constraints.
    \item A case study demonstrating the achievable performance improvement of allowing non-static traffic distributions for a given junction infrastructure.
\end{enumerate}

The rest of the paper is organized as follows: In Section~\ref{sec:rw}, we begin with a brief review of related work, followed by background on the methodology in Section~\ref{sec:background}.
Afterwards, Section~\ref{sec:problem} presents the traffic rate assignment problem formulation. The queueing-based model for the non-linear black-box constraints is described in Section~\ref{sec:qs_all}, followed by our BO solution approach in Section~\ref{sec:bo}. 
Section~\ref{sec:results} demonstrates the efficacy of our method, and Section~\ref{sec:discussion} concludes the work.

\section{Related Work}\label{sec:rw}

In order to determine the performance capability of their network, infrastructure managers can choose from a variety of methods.
For example, the UIC Code 406~\cite{uic406_2013} defines the metric of \textit{capacity utilization} for railway lines (and nodes) with an occupancy time rate as the quotient of summed occupancy times in a given timetable and the considered time period. 
For railway lines and junctions, this can be calculated by compressing the blocking times of all subsequent train runs within the section~\cite{uic406_2013, landexMeasuresTrackComplexity2013, goverdeRailwayLineCapacity2013}.
By formulating the occupation parameters of a given timetable in Max-Plus-Algebras, capacity utilization may also be computed for railway networks \cite{goverdeRailwayTimetableStability2007, besinovicCapacityAssessmentRailway2018}.
However, this occupation ratio is strongly dependent on the sequence of train runs and therefore the given timetable.
Even though methods to efficiently sample possible timetable requests to achieve independence from the timetable itself exist~\cite{jensenStrategicAssessmentCapacity2017, jensenDeterminationInfrastructureCapacity2020, weikExtendingUIC406based2020}, 
other capacity metrics have been considered as well.
These metrics can be partitioned by their capacity understanding,  \cite{jensenStrategicAssessmentCapacity2017, emundsEvaluatingRailwayJunction2024} distinguish between \textit{(i) theoretical capacity},
describing the theoretical maximum number of trains on an infrastructure, satisfying mainly safety and driving dynamic constraints; 
\textit{(ii) timetable capacity}, corresponding to the maximum number of requests in a timetable for the given infrastructure, additionally considering operating program specific settings and quality thresholds; and \textit{(iii) operational capacity}, the maximum number of trains with acceptable operational quality, also respecting disturbances and delay propagation.

Theoretical capacity analysis is mainly utilizing mixed integer programming (MIP) formulations, saturating a given timetable with additional train requests \cite{burdettTechniquesAbsoluteCapacity2006, liaoRailwayCapacityEstimation2021} or independently allocating occupation slots to requested train journeys either for capacity determination directly \cite{yaghiniINTEGERPROGRAMMINGMODEL2014, burdettMultiobjectiveModelsTechniques2015} or within timetabling problems \cite{caimiNewResourceConstrainedMulticommodity2011, leutwilerLogicbasedBendersDecomposition2022}.
However these methods typically consider mainly conflict-freeness and refrain from including quality concerns.
In order to assess the operational quality, infrastructure managers might decide to analyze operational data directly \cite{spanningerTrainDelayPredictions2023, cormanEstimatingAggregateRailway2022}, utilize simulation approaches \cite{abrilAssessmentRailwayCapacity2008,schmidtErhohungEffektivitatUnd2010, liangIncreasingPerformanceRailway2017} or perform analytical analysis of delay propagation \cite{schwanhausserBemessungPufferzeitenIm1974, bukerStochasticModellingDelay2012, sahinMarkovChainModel2017}.

Other analytical approaches focus on determining the timetable capacity of railway lines \cite{Wendler.2007, weikLongtermCapacityPlanning2020}, junctions \cite{nilsniessenWaitingLossProbabilities2013, schmitz2017markov}, stations  \cite{bychkovModelingRailwayStations2021, Emunds_Station}, or networks \cite{huismanSolvableQueueingNetwork2002a, kazakovApproachRailwayNetwork2023}, modeling the timetabling process with queueing systems.
In particular, we build on the approach from \cite{emundsEvaluatingRailwayJunction2024} which calculates the timetable capacity of railway junction infrastructure by formulating a continuous-time Markov chain (CTMC) and evaluating it with probabilistic model-checking.

Existing approaches to integrate dynamic traffic distributions into performance estimations for railway infrastructure are either limited to linear approximations of gridwise predetermined capacity results \cite{Emunds_Capa2MIP}, or calculate theoretical capacity, \eg by directly including timetabling problems into line planning formulations \cite{fuchsEnhancingInteractionRailway2022}, which does not consider timetable quality indicators.

In order to include dynamic traffic distributions, this paper introduces an approach to iteratively enhance the assignment of route frequencies, obtaining queue-length estimations for the selected assignment in each step.
This process necessitates numerous calls to the model-checker to evaluate constraints partially described by CTMCs.
To limit the amount of required iterations, we leverage BO for an efficient optimization under expensive to evaluate constraints. 

BO increased in popularity for optimizing unknown objective function in a data-efficient manner with applications ranging from robotics \cite{calandra2016bayesian,berkenkamp2016safe,neumann2019data} to drug discovery \cite{griffiths2020constrained,colliandre2023bayesian,maus2022local}.
In railway systems, BO has been applied to network optimization~\cite{hickish2020investigating}, train delay prediction modelings~\cite{shi2020train,luo2023multi}, and interval dynamic analysis~\cite{wan2020new}.
BO has not been applied to optimizing railway junction performance.

A key factor contributing to the popularity of BO is the flexibility of its framework.
It has been extended, \eg to high-dimensional settings~\cite{kirschner2019adaptive,erikssonScalableGlobalOptimization2020,muller2021local}, time-varying settings~\cite{bogunovic2016time,bardou2024too,brunzemaevent}, as well as optimizing under unknown constraints~\cite{gardner2014bayesian,gelbart2014bayesian,eriksson2021scalable}.
For a comprehensive overview, we refer to \cite{garnett2023bayesian}.
Furthermore, the GP surrogate model used in BO can directly incorporate prior knowledge about the underlying problem by, \eg designing task specific kernels \cite{marco2017design} or mean functions.
In this paper, we build on BO with unknown constraints and discuss the specific case where the objective function is known.
To our knowledge, this work represents the first explicit treatment of BO scenarios with known objectives but black-box constraints.
Furthermore, we demonstrate that performance can be significantly improved by incorporating a problem-specific mean function for modeling the black-box constraints.


\section{Background and Preliminaries}\label{sec:background}

This paper combines BO with queuing-based performance estimation methods to obtain solutions for traffic rate assignment problems.
In this Section, we recall the performance metric \textit{timetable capacity} 
(Section \ref{sec:tt_capa}) along with its associated quality thresholds 
(Section \ref{sec:thresholds}), which will be used to formulate the constraints on our optimization problem.
Additionally, we revisit fundamentals of GPs (Section \ref{sec:gp}) and BO (Section \ref{subsec:bo}).
All notation is summarized in the \ref{app:vars} in Table~\ref{tab:variables} to facilitate parsing and to serve as a reference throughout the paper.

\subsection{Timetable Capacity of Railway Systems}
\label{sec:tt_capa}

Successful railway transportation of goods and passengers depends on critical long-term planning processes that occur several years before operations begin.
To ensure adequate design of costly infrastructure, evaluation must be conducted independently of specific timetables, which may change throughout the infrastructure's multi-decade lifespan.

This work employs the performance metric of \textit{timetable capacity} (see also \cite{Wendler.2007, emundsEvaluatingRailwayJunction2024}).
It corresponds to the capacity of the railway infrastructure in the timetabling process, \ie the assignment of conflict-free occupation slots to requested train journeys.
Since multiple train journeys might request the occupation of conflicting infrastructure routes at the same time, the infrastructure manager needs to reschedule some requests to later occupation times.
Given a route~$r$, the number of rescheduled requests awaiting their corresponding occupation slot can be represented as a queue for any point in time.
Note that the average queue length of occupation requests depends on their stochastic arrival and service processes.
In this work, we utilize 
continuous-time Markov chains
to obtain an approximation of the expected queue length $L_r$ for every route $r$ (see Section \ref{subsec:ql_estimations}).
These average queue lengths on all routes can be utilized as performance metrics for the infrastructure of the railway junction by comparing them to thresholds for adequate quality.

\subsection{Thresholds on Queue Lengths}
\label{sec:thresholds}

To qualitatively evaluate performance capabilities, different thresholds have been established for corresponding capacity metrics.
For example, maximum occupation rates have been defined \cite{uic406_2013} for the use with timetable compression methods.
Following the methods of the largest European infrastructure manager, DB InfraGO \cite{dbnetzagRichtlinieFahrwegkapazitat2009}, we use a threshold for the expected average queue length.
Developed in a survey \cite{schwanhausserErmittlungQualitatsmassstabenFur1982} under railway dispatchers to evaluate timetable quality on a railway line, 
the limit on the average queue length is defined as 
\begin{equation}
\label{wsl_limit_sh}
  L_{\text{limit, r}} = 0.479 \cdot \mathrm{exp}(-1.3 \cdot p_{\text{pt}, r}) ~,
\end{equation}
in dependence of the proportion of passenger traffic $p_{\text{pt}, r}$.
Note that this threshold is defined for railway lines and we hence apply it independently for every route within a railway junction.

\subsection{Basics on Gaussian Processes}\label{sec:gp}

We will use GPs \cite{rasmussen2006gaussian} to model the the black-box constraints formulated in the previous subsection.
A GP is fully defined by a mean function $m \colon \feasibleset \to \R$ and a kernel function $k \colon \feasibleset \times \feasibleset \to \R$ and we will denote it as $\mathcal{GP}(m(\optvar), k(\optvar,\optvar'))$.
This prior GP can be conditioned on a dataset $\mathcal{D}_t\coloneqq\left\{(\optvar_i,y_i)\right\}_{i=1}^{t-1}$ and evaluated on a test point $\optvar$ yielding a normal predictive distribution $\mathcal{N}(m_{\mathcal{D}_{t}}(\optvar), \sigma^2_{\mathcal{D}_{t}}(\optvar))$ with mean and variance as
\begin{align}
    m_{\mathcal{D}_{t}}(\optvar) &= m(\optvar) - \bm{k}_t(\optvar)^\top \left( \textbf{K}_t + \sigma_n^2 \textbf{I} \right)^{-1} (\bm{y}_t -m(\optvar)) \\
    \sigma^2_{\mathcal{D}_{t}}(\optvar) &= k(\optvar,\optvar) - \bm{k}_t(\optvar)^\top \left( \textbf{K}_t + \sigma_n^2 \textbf{I} \right)^{-1} \bm{k}_t(\optvar). \label{eq:gp}
\end{align}
Here, $\textbf{K}_t = [k(\optvar_i, \optvar_j)]_{i, j = 1}^{t-1}$ is the Gram matrix, $\bm{k}_t(\optvar)= [k(\optvar_i, \optvar)]_{i=1}^{t-1}$ is a vector, and $\bm{y}_t=[y_1, \dots, y_{t-1}]^\top$ are noisy measurements.
A key advantage of GPs is that their mean and kernel functions can encode prior knowledge about the unknown function's behavior, which we leverage to enhance constraint expressiveness and overall performance.

\subsection{Overview on Bayesian Optimization}\label{subsec:bo}
Because they can yield well-calibrated uncertainty estimates, GPs are widely adopted as surrogate models within BO \cite{garnett2023bayesian}.
As a sample-efficient global optimization method for black-box objective functions, BO is especially suited for problems where function evaluations are expensive.
In its standard form, BO aims to find $\optvar^* = \arg\max_{\optvar \in \feasibleset} f(\optvar)$ by sequentially querying the unknown objective $f(\optvar)$ and receiving zeroth-order feedback, \ie only the function value and not its gradient.
Here, $\feasibleset \subseteq \R^d$ is the feasible set where $d$ is the dimension of decision variable.
As stated above, typically, $f(\optvar)$ is modeled as a GP and the next query point $\optvar_t$ at each iteration $t \in \mathcal{I}_T \coloneqq \{1,\dots, T\} $ is selected by optimizing an acquisition function $\alpha : \feasibleset \to \mathbb{R}$ as
\begin{equation}
    \optvar_{t+1} \gets \arg\max_{\optvar \in \feasibleset} \alpha(\optvar \mid \mathcal{D}_t).
\end{equation}
This acquisition function usually balances exploration (sampling in uncertain regions) and exploitation (sampling near well-performing regions) to efficiently converge to the optimum. 
There exist various choices for an acquisition function such as upper confidence bound \cite{auer2002using, srinivas2010gaussian}, expected improvement \cite{jones1998efficient}, and Thompson sampling \cite{thompson1933likelihood}.
In this paper, we will consider variants of the latter two acquisition functions but we will tailor them to our specific setting of a known objective function under black-box constraints.


\section{Problem Definition}\label{sec:problem}

One major long-term planning task is to assign frequencies to paths within a railway network, in order to best satisfy a given demand of traffic for origin-destinations combinations.
This problem can be formulated for large-scale transportation networks and even include considerations regarding different vehicle-sizes or passenger connections between the various lines.
The focus of this work, however, is not to solve such a \textit{line-planning} problem for a complex railway network, but rather on introducing a novel method to model capacity restrictions for railway junctions, connecting the lines between stations within such networks.
This allows for assessing infrastructure performance indicators with non-static traffic distributions, enhancing the level of flexibility included into applications for long-term planning problems.



We model a railway junction $J=\left(\Routes,C\right)$ with a set of $k$ routes $\Routes$ and a \textit{conflict matrix} $C \in \{0,1\}^{k \times k}$, describing whether the parallel occupation of two routes $r, r^{\prime}$ is permitted ($C_{r, r^{\prime}} = 0$).
One major influence to the occupation time of railway infrastructure is the rolling stock, we therefore notate an occupation request $o \in \Occs$ as a tuple $o=(r, u) \in \Routes \times \mathcal{U}$ of a selected route $r$ and a corresponding rolling stock unit (train type) $u \in \mathcal{U}$.
We sometimes refer to $C_{o, o^{\prime}} := C_{r, r^{\prime}}$ to model conflicts for a combination of requests $(o, o^{\prime}) = ((r, u), (r^{\prime}, u^{\prime}))$.

In this work, we measure the timetable capacity of a railway junction within a fixed time horizon $t_U$.
The number (or \textit{frequency}) $f_o$ of trains per request $o \in \Occs$ can be modeled as an \textit{arrival rate}
\begin{equation}
    \ovar_o = \frac{f_o}{t_U},
\end{equation}
describing the number of trains of request $o$, arriving at the modeled junction in the time horizon $t_U$.
The number of different requests is denoted with $d = |\Occs|$, describing the dimension of the assignment vector $ (\ovar_o)_{o \in \Occs} = \ovar \in \R^d$.

For a given upper bound function $\text{ub}: \Occs \rightarrow \mathbb{R}, o \mapsto \text{ub}_o $, we can formulate the
constraints 
\begin{align}
    &\ovar_{o} \leq \text{ub}_{o}, \quad\,\, \forall  o \in \Occs ~, \label{equ:mip_nomorethandemand}\\
   & c_r \left(\ovar\right) \leq  0 , \quad \forall  r \in \Routes~, \label{equ:mip_capacity} 
\end{align}
describing the feasible set of $\ovar$.
In this work, the \textit{capacity restriction} functions $c_r: \mathbb{R}^{d} \rightarrow \mathbb{R}$ are utilized to include limitations arising from the timetable-independent capacity metric (see Section~\ref{sec:tt_capa}) computed via queuing systems for the railway junction infrastructure.
\footnote{
Note that the core ideas of this paper are not restricted to employing a queuing-based capacity model.
However, we will tailor parts of our method in Section~\ref{sec:bo}) to this modeling approach and exploit the exponential behavior of the capacity constraints for improved performance.}


When determining the performance capability of a railway junction, the intuitive objective function
\(
    \max_{\ovar} \sum_{o} \ovar_{o},
\)
might give the theoretical maximum of traffic rate assignments.
However, the resulting rates might not represent an optimal assignment in practice, since neither the traffic nor the capacity consumption are distributed equally between all combinations.
Therefore, it is necessary to extend the objective function with terms representing desired traffic parameters.

We utilize two different approaches to tackle the practicality of obtained solutions in this work.
In detail, the deviation from two different distribution targets, \ie the spreading $\tilde{p}_u$ (or $\tilde{p}_r$) between train types $u\in \mathcal{U}$ or routes $r \in \Routes$, are discussed.
Naturally, a chosen traffic rate assignment $\ovar$ also defines the corresponding distributions between train types $p_u(\ovar)$ and routes $p_r(\ovar)$.
An objective formula of the form
\begin{equation}
    \max_{\ovar} \sum_{o \in \Occs} \ovar_o   - w_U \cdot \sum_{u \in U}  \left( p_u (\ovar) - \tilde{p}_u\right)^2  \label{eq:obj_general},
\end{equation}
 can be utilized to enforce adherence to a given train type distribution $\left(\tilde{p}_u\right)_{u \in \mathcal{U}}$.
The objective \eqref{eq:obj_general} incorporates a penalty, corresponding to the summed squared differences $\sum_{u \in \mathcal{U}}  \left( p_u - \tilde{p}_u\right)^2$, into the objective value by weighting it with a weighting factor $w_U \in \mathbb{R}$.
An analogous objective formula can be developed for distributions across other traffic contexts, such as the different train routes, or combinations of these parameters.
With this, we can formulate the \textit{Penalized Traffic Rate Assignment Problem}~(\OPT) that we aim to solve in this paper.

\begin{tcolorbox}[noboxblank,title={Penalized Traffic Rate Assignment Problem \hfill {\normalfont (\OPT)}}, label=OPT]
\begin{align}
    \ovar^* = \argmax_{\ovar} \quad &\sum_{o \in \Occs} \ovar_o - w_U \cdot \sum_{u \in \mathcal{U}}  \left( p_u (\ovar) - \tilde{p}_u\right)^2 \tag{a}\label{eq:OPT-a} \\
    \text{subject to} \quad &\ovar_{o} \leq \text{ub}_{o} \qquad\,\, \forall o \in \Occs \tag{b}\label{eq:OPT-b} \\
    &c_r \left(\ovar\right) \leq 0 \qquad \forall r \in \Routes \tag{c}\label{eq:OPT-c}
\end{align}
\end{tcolorbox}

\section{Queuing-Based Capacity Restrictions}
\label{sec:qs_all}
After deriving \OPT, the remaining step is to formalize the capacity restriction functions $c_r \left(\ovar\right)$ in \eqrefOPT{eq:OPT-c}.
For this, Section~\ref{sec:qb_capa} introduces a model of queueing systems for railway infrastructure dimensioning along with the technique to obtain queue-length estimations for railway junctions in Section~\ref{subsec:ql_estimations}.
In a final step in Section~\ref{subsec:constr_formulation}, we obtain constraint formulations for the traffic rate assignment problem \OPT considered in this work.

\subsection{Queueing Systems for Railway Performance Measures}
\label{sec:qb_capa}

The timetable capacity of a railway junction $J = \left(\Routes,C\right)$ not only depends on the layout of the infrastructure $\Routes,C$ and the utilized rolling stock units $\mathcal{U}$, but additionally on the distribution of traffic to different routes.
A railway junction might be traversed with heterogeneous traffic types, such as passenger and freight traffic, all of which utilizing different rolling stock units. 
Therefore, the occupation time $b: \Occs \rightarrow \mathbb{R}, o \mapsto b_o$ of a request $o= (r, u)$ depends on its route $r$ and rolling stock unit $u$.

Furthermore, the occupation time of a request within a railway junction is not only influenced by the train itself, but additionally by the following train.
In railway operations research, the occupation time is therefore usually modeled as a minimum headway time $h: \Occs \times \Occs \rightarrow \mathbb{R}, (o, o^{\prime}) \mapsto h_{o, o^{\prime}}$ for a sequence of trains $(o, o^{\prime})$, describing the minimum time-span between the start of the occupation of the first train $o$, until the start of the occupation of the following train $o^{\prime}$. 
This minimum headway time is calculated once for every possible request combination on microscopic infrastructure models, more details regarding the blocking-time model and the computation of minimum headway times can, \eg be found in \cite{hansenRailwayTimetablingOperations2014}.

For timetable-independent planning problems, the sequence of trains is not predetermined. 
Hence, the occupation time $b_o$ needs to be formulated without knowledge of following requests.
We therefore assume a uniform distribution of request sequences and describe the average occupation time 
\begin{equation}
\label{eq:oc_times_req}
    b_o =  \frac{\sum_{\substack{o^{\prime} \in \Occs \\ C_{o, o^{\prime}} = 1}} \ovar_{o^{\prime}} \cdot h_{o,o^{\prime}}}{\sum_{\substack{o^{\prime} \in \Occs \\ C_{o, o^{\prime}} = 1}} \ovar_{o^{\prime}}}
\end{equation}
of a single request $o$ by weighting the minimum headway times $h_{o, o^{\prime}}$ of conflicting request combinations $o, o^{\prime}$ with their arrival rates.

In this work, we model the timetabling process on the railway junction as a queuing system (see Section \ref{subsec:ql_estimations}).
Within this queueing system, traffic on the junction $J=\left(\Routes,C\right)$ is decomposed along the routes $r \in \Routes$, and performance indicators are calculated for every route.
Hence, we need to abstract the occupation times from the request based formula in \eqref{eq:oc_times_req} to the average occupation time
\begin{equation}
\label{eq:oc_time_r}
    b_r = \frac{\sum_{\substack{o=(r,u) \in \Occs}} \ovar_o \cdot b_o}{\lambda_r} 
\end{equation}
for a route $r$ by weighting the request occupations times $b_o$ for all requests $o = (r,u) \in \Occs$ on the route $r$ with their respective arrival rates. In \eqref{eq:oc_time_r}, $\lambda_r = \sum_{(r,u) \in \Occs} \ovar_{(r,u)}$ defines the total arrival rate to a route $r$.
The with \eqref{eq:oc_time_r} obtained route-based occupation times $b_r$ allow us to define the \textit{service rate} $\mu_r = 1/b_r$ for a route $r$, giving the average rate with which the requests are serviced within the queuing system.
These service rates $\mu_r$ can be compared to the arrival rates $\lambda_r$, in the \textit{occupation ratio} $\rho_r = \lambda_r/\mu_r$, serving as an indicator of the utilization of a route $r$.

In the modeled queuing system, arrived but not yet started requests are gathered in an individual queue for every route. The expected number of request in the queue of route $r$, denoted as the \textit{expected queue-length} $L_r$, gives another performance indicator of the queuing system.
The next Section will describe the utilized approach to calculating these $L_r$ from obtained arrival and service rates.

\subsection{Obtaining Queue-Length Estimations}
\label{subsec:ql_estimations}

The queueing system for the railway junction with arrival and service rates derived in Section \ref{sec:qb_capa} can be formulated as a CTMC with state space $S$, see \cite{emundsEvaluatingRailwayJunction2024} for full details.

To compute the expected queue-length $L_r$ for each route $r$ in the modeled railway junction, 
the stationary distribution of the CTMC can be analyzed. 
The stationary distribution is a probability distribution over the state set $S$, i.e., 
it maps a probability $p(u)$ to each state $u \in S$, indicating the likelihood that 
the system will be in the corresponding state in the long run.

Such a steady state analysis has been used by a number of analytical performance determination approaches, see f.e. \cite{Wendler.2007, nilsniessenWaitingLossProbabilities2013, weikLongtermCapacityPlanning2020}.
In this work, we build on an approach introduced in \cite{emundsEvaluatingRailwayJunction2024} which builds the CTMC in the formal PRISM language 
\cite{DaveParkerGethinNormanMartaKwiatkowska.2000} and obtains the queue-length estimations $L_r$ with probabilistic model-checking \cite{Hensel.2022}.
This approach relies on an approximation of the queueing system by limiting the number of waiting positions in the queue of each route $r$ with a finite number $B \in \mathbb{N}$.
We use values of $B=3$ and $B=5$ for the computations in Section \ref{sec:results}, resulting in the loss of every request arriving to a route with three (five) waiting requests, therefore under-approximating the theoretical average queue-length. 
Further influences of this parameter $B$ are discussed in \cite{emundsEvaluatingRailwayJunction2024}.

Note that due to the Markov property of the CTMC, resulting estimations are obtained for exponentially distributed inter-arrival and service times only ($M/M$ systems \cite{kendallStochasticProcessesOccurring1953}).
To adapt this system to other probability distributions ($GI/GI$ systems \cite{kendallStochasticProcessesOccurring1953}), we can utilize the formula of Hertel \cite{hertel1984exakte}, where
\(
\label{approx_ELW}
    L_{r}(M/M) \cdot \frac{1}{\gamma} \approx L_{r}(GI/GI)
\)
with parameters
\(
\label{approx_ELW_gamma}
    \gamma = \frac{2}{c \cdot v_S^2 + v_A^2}
\)
and 
\(
\label{approx_ELW_c}
    {c = \left(\frac{\rho_r}{s}\right)^{1-v_A^2} \cdot (1+v_A^2) -v_A^2~},
\)
and scale the queue-lengths $L_r$ accordingly, using coefficients of variation for the service and arrival process.
In this work, we scale the queue-length with a variation coefficient of the arrival of $v_A = 0.8$ and of the service process of $v_S = 0.3$, according to standard values.
The next section highlights a formulation of the capacity constraints by using the obtained queue-length estimates.

\subsection{Capacity Constraint Formulation}
\label{subsec:constr_formulation}

After obtaining the estimated queue-lengths $L_r$ for every route, we can formulate the constraints for the traffic rate assignment problem from Section~\ref{sec:problem}.
For this, constraint \eqrefOPT{eq:OPT-c} can be updated with the comparison of the in Section~\ref{subsec:ql_estimations} estimated queue-length $L_r=L_r\left(\ovar\right)$ for a given rate assignment $\ovar$ with the threshold value $L_{\text{limit, r}}$ from Section \ref{sec:thresholds} to the constraint
\begin{equation}
    \text{\eqrefOPT{eq:OPT-c}:} \quad c_r \left(\ovar\right) = L_r\left(\ovar\right) - L_{\text{limit, r}} \leq  0 , \, \forall  r \in \Routes~, \label{equ:mip_capacity_final}
\end{equation}
for every route $r$.


In practice, estimations of $L_r\left(\ovar\right)$ can be expensive to compute and are nonlinear.
To still efficiently solve \OPT, we approximate the constraints \eqrefOPT{eq:OPT-c} with GPs and leverage BO for an informed search.

\section{Utilizing Bayesian Optimization}
\label{sec:bo}

We next tailor standard BO approaches to our optimization problem at hand, \ie a known objective function and black-box constraints.
This differs from the standard BO formulation where the objective is usually unknown. 
In this section, we will include the additional information about the objective and its unconstrained optimum into the design of our method.
We begin by describing how to model the black-box constraints.

\subsection{Modeling Unknown Constraints on the Capacity}
\label{subsec:bo_constr}

Since the constraints $c_r (\optvar)$ are expensive to evaluate, we treat them as unknown and model them with a GP. 

\subsubsection{Kernel function}
In this work, we will choose as the kernel function a Mat\'ern kernel \cite{rasmussen2006gaussian} with $\nu = 5/2$ and individual lengthscales $\ell_i$, scaling the correlation between inputs for all dimensions, as
\begin{equation}
    k(\optvar, \optvar') = \sigma_f^2 \left( 1 + \sqrt{5} \Delta + \frac{5}{3} \Delta^2 \right) \exp \left( -\sqrt{5} \Delta \right),
\end{equation}
where $\sigma_f^2$ is the output scale, and $\Delta \coloneqq \sqrt{\sum_{o=1}^d \left[(\optvar_o - \optvar'_o)^2 / \ell_o^2 \right]}$ is the normalized Euclidean distance.

\subsubsection{Mean function}
\begin{wrapfigure}{r}{0.55\textwidth}
    \vspace{-1.2em} 
    \includegraphics[width=0.52\textwidth]{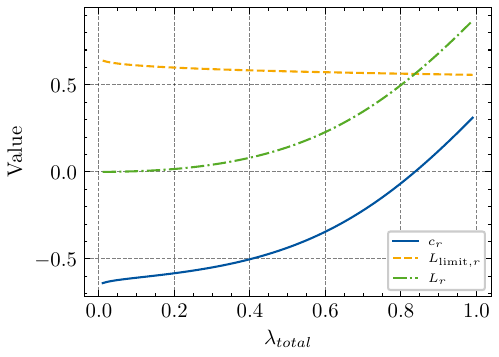}\vspace{-0.5em}
    \caption{Exemplary plot of a constraint $c_r$ and its components $L_r, L_{r, \text{limit}}$ for different traffic rates $\lambda_{\text{total}}$.
    }
    \label{fig:intuition_constraint}
    \vspace{-1em}
\end{wrapfigure}

For the mean function, we include intuition on the shape of the constraints.
For this, we visualize the constraint for a one-dimensional problem on a dense grid as shown in 
Fig.~\ref{fig:intuition_constraint}. 

Observing an exponential trend in the constraint $c_r$,
we leverage the fact that embedding prior knowledge into the GP model is possible and formulate the following mean function for our GP prior of $c_r$ as
\begin{equation}
    m_{\text{exp}}(\optvar) = \beta \exp\left(\sum_{i=1}^d w_i \optvar_i\right) - \gamma \, . \label{eq:exp_mean}
\end{equation}
Here, $ \beta$, $ w_i $, and $ \gamma $ are learnable parameters.
To ensure the observed positive trend, we enforce positivity on $ \beta $ and $ w_i $ by re-parameterizing them as
\begin{equation}
    \beta = \exp(\beta_r) \text{ and } w_i = \exp(w_{i,r}) \,\, \forall i \in \mathcal{I}_d
\end{equation}
where $ \beta_r\in \R$ and $ w_{i,r} \in \R$ are the actual learnable parameters in an unconstrained space.
In our experiments, we will also compare this to the standard approach of using a constant mean function.
Importantly, the formulation in \eqref{eq:exp_mean} defines an abstraction that naturally encompasses the constant mean function as a special case.
Specifically, when $\beta \to 0$ or $w_i \to 0$ for all $i$, the mean function simplifies to a constant, making it a subset of our proposed mean's solution space.
This flexibility enables the model to recover a simple constant offset when exponential patterns are absent, while capturing exponential trends when they are present in the data.
To summarize, the set of learnable parameters for each constraint is $\Theta_r = \{\sigma_n, \ell_0, \dots, \ell_d, \sigma, \beta_r, w_0, \dots, w_d, \gamma_r \} \in \R^{2d+4}$.
We will learn these parameters at each iteration given all data set by maximizing the marginal log-likelihood~\cite{rasmussen2006gaussian}.

\subsection{Choosing the Next Query with Known Objectives under Unknown Constraints}
\label{subsec:acqui_funcs}

We next discuss how to design an acquisition function for our specific problem formulation.
Here, we will evaluate two approaches: \textit{i)} reformulating constrained expected improvement to a known objective and \textit{ii)} using Thompson sampling.

\subsubsection{Adaptive Trust Regions}\label{sec:tr}
For both approaches, we will constrain the feasible set to a trust region as in the algorithm TuRBO proposed in \cite{erikssonScalableGlobalOptimization2020}.
TuRBO maintains a trust region $\mathcal{TR}_t$ as the hyperbox around parameters of the best observed value $\hat{\optvar}^*_t$ based on the lengthscales of the GP surrogate of the objective.
Crucially, in our setting, the objective is known.
Therefore, we also omit the scaling and define the trust region
\begin{align}
    \mathcal{TR}_t = &\left\{ \optvar \in \feasibleset \Bigm| \vert \optvar_j - \hat{\optvar}^*_{t,j} \vert \leq \frac{L_t}{2}, \, \forall i \in \mathcal{I}_d \right\}.
\end{align}
Here, $L_t$ is the hyperbox length at the current iteration.
We adapt the this parameter over time in the same fashion as in \cite{erikssonScalableGlobalOptimization2020}, \ie we shrink it if we have not improved in the past few iterations and expand it if improved in successive iterations.
We use the same hyperparameter as \cite{erikssonScalableGlobalOptimization2020} to which we refer to for further details.

\subsubsection{Adapting Constrained Expected Improvement to Known Objective}
With knowledge of the objective function, we no-longer have to consider the expected improvement (EI) of a next query compared to the previous one as in the standard formulation, as the expectation is taken with respect to a fixed objective. 
Additionally, the improvement over the best query is merely a shift in the function values and does not impact its optimizer.
Therefore, we can directly reformulate the classic EI acquisition function as
\begin{equation}\label{eq:norm_aq}
    \alpha_{\text{CI}}(\optvar) \coloneqq f(\optvar) \cdot \mathbb{P}(c_r(\optvar) \leq 0) 
\end{equation}
where the feasibility probability is given by
\begin{equation}
    \mathbb{P}(c_r(\optvar) \leq 0) = \prod_{r\in \Routes}\Phi \left(\frac{-m_r(\optvar)}{\sigma_r(\optvar)} \right),
\end{equation}
and $ \Phi(\cdot) $ is the cumulative distribution function of the standard normal distribution.

To ensure numerical stability, we follow advise from \cite{ament2023unexpected} and instead of optimizing \eqref{eq:norm_aq}, we optimize a logarithmic version for numerical stability as
\begin{equation}
    \optvar_{t+1} \gets \argmax_{\optvar \in \mathcal{TR}_t}  \,\,\log f(\optvar) + \sum_{r\in \Routes} \log \mathbb{P}(c_r(\optvar) \leq 0).
\end{equation}
Interestingly, this formulation closely resembles a log-barrier approach commonly used in classical numerical optimization \cite{nocedal1999numerical}.
In traditional log-barrier methods, constraints are enforced by introducing penalty terms of the form $-\sum_{j} \log (-c_r(\optvar))$.
As the optimization progresses, these barrier terms encourage feasible solutions while smoothly guiding the optimizer towards the constraint boundary.
Similarly, as our model of the constraint improves, the acquisition function will increasingly favor points closer to the active constraints.


\begin{wrapfigure}{r}{0.55\textwidth}
    \centering
    \vspace{-0em} 
    \includegraphics[width=0.5\textwidth]{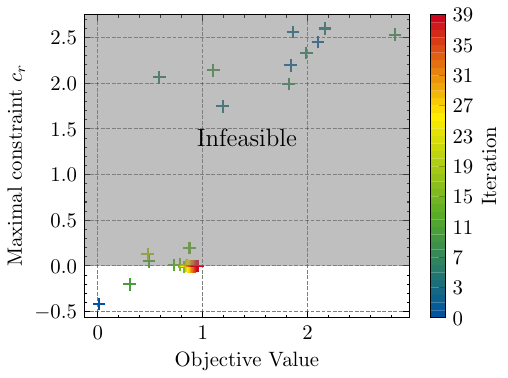}
    \caption{Iterations of \texttt{EI-Exp-TR} (see Section~\ref{sec:model_select}) and their corresponding maximal constraint $\max c_r$ and objective value.}
    \label{fig:feasibility_turbo}
    \vspace{-2em}
\end{wrapfigure}

In Fig~\ref{fig:feasibility_turbo}, we show a small example for the applicability of our revised mean function in combination with an adaptive trust region and our EI formulation.
We can observe that over time, the algorithm will converge to a feasible solution respecting our unknown constraints.

\subsubsection{Thompson Sampling in Constrained Bayesian Optimization}

Besides an EI-like acquisition function, we also apply constrained Thompson sampling (TS) to solve the \OPT.
The core idea of TS is to choose the next query location as the global optimizer of a posterior sample from the surrogate model.
Here, we directly follow a constrained version of TS proposed in \cite{eriksson2021scalable} but adapt it to the setting of a known objective function.
For this, we select $M$ candidates $\mathcal{M} = \{\optvar^{(0)}, \dots, \optvar^{(M)}\} \subseteq \mathcal{TR}_t$ from a pseudo random Sobol sequence from the current trust region.
We next sample realizations from the posteriors distributions of all the constraints and select the query location for the next iteration as
\begin{equation}\label{eq:best_point}
    \optvar_{t+1} \gets
    \begin{cases}
        \displaystyle\argmin_{\optvar\in \mathcal{F}} f(\optvar),                                      & \text{if } \mathcal{F} \neq \emptyset, \\
        \displaystyle\argmin_{1 \leq j \leq M} \sum_{r \in \Routes} \max\left(0, c_r\left(\optvar^{(j)}\right)\right), & \text{otherwise},
    \end{cases}
\end{equation}
where $\mathcal{F} = \left\{ \optvar \in \mathcal{M} \,\middle| \, c_r\left(\optvar\right) \leq 0, \,\forall r \in \Routes \right\}$ denotes the set of feasible points among the $M$ candidates.

\subsection{Recenter Trust Region on Probable Optimum}
In addition to the conventional TuRBO framework, which iteratively establishes a trust region 
centered around the highest previously known objective value to bound the optimization area for the subsequent optimization, 
we propose an alternative method for determining the trust region.
With this approach, we aim to leverage the availability of an analytical description of the objective function.

In detail, the analytical description of the objective function 
\eqrefOPT{eq:OPT-a}
as a composition of differentiable functions allows us to determine the gradient analytically.
Under the assumption that constraints \eqrefOPT{eq:OPT-b} and \eqrefOPT{eq:OPT-c} are satisfied within a region $\mathcal{F}_{\feasset} \subseteq \mathbb{R}^d$, we can determine a local maximum $\ovar^{\ast}_t = \argmax_{\ovar \in \mathcal{F}_{\feasset}} \text{\eqrefOPT{eq:OPT-a}}$ using non-linear optimization methods such as sequential least-squares programming \cite{kraft1988software}.


However, with the probabilistic representation of the constraints, defining a feasibility region $\mathcal{F}_{\feasset}$ involves managing uncertainties.
Specifically, we aim to solve 
the non-linear chance constrained optimization problem
\begin{equation}
    \begin{aligned}
        \ovar^{\ast}_t = \argmax_{\ovar \in \feasset}~~ & \text{\eqrefOPT{eq:OPT-a}} &\\
    \text{subject to}~~ & \text{\eqrefOPT{eq:OPT-b}}\\
     & \mathbb{P}\{ c_r(\ovar)\leq 0\} \leq \alpha_r,& \forall r\in \Routes~.
    \end{aligned}\label{eq:sub}
\end{equation}
Given that $c_r(\ovar_{o})$ is Gaussian distributed, we can state that
\begin{equation}
\mathbb{P}\{ c_r(\ovar)\leq 0\} \leq \alpha_r \Leftrightarrow  m_{\mathcal{D}_t, r}(\ovar) + \Phi^{-1}(\alpha_r)\, \sigma_{\mathcal{D}_t, r}(\ovar) \leq 0
\end{equation}
with $\Phi^{-1}(\cdot)$ as the inverse CDF
which again yields a tractable optimization problem.
We utilize its optimal solution $\ovar^{\ast}_t$ as the center of a new optimization region $\mathcal{TR}_{t}^{\prime} \subseteq \mathbb{R}^d$
to be used with the acquisition functions discussed in Section \ref{subsec:acqui_funcs}.
By formulating \eqref{eq:sub} as a chance constrained problem, we ensure feasibility of the center $\ovar^{\ast}_t$ of $\mathcal{TR}_{t}^{\prime}$ with high probability.\footnote{In all following example, we directly choose $\Phi^{-1}(\alpha_r) = 2$ which roughly corresponds to $\alpha_r \approx 0.975$, \ie $97.5 \%$ satisfaction probability.}

In this work, we compute the gradient with SymPy~\cite{SymPy}, and use SLSQP~\cite{kraft1988software} as implemented in SciPy~\cite{2020SciPy-NMeth}
to find the maximizer $\ovar^{\ast}_t$.
Subsequently, the next optimization region $\mathcal{TR}_{t}^{\prime}\subseteq \mathbb{R}^d$ is defined as a hypercube centered around $\ovar^{\ast}_t$, with an iteratively adjusted edge length, similar to the approach used in the TuRBO framework (\cf Section~\ref{sec:tr}).

\section{Results}\label{sec:results}

We next benchmark our approaches.
First, we evaluate different combinations of our three main proposed adaptations on a small exemplary railway network (Section~\ref{sec:model_select}). 
These adaptations comprise: \textit{(i)} the inclusion of prior knowledge in the GP model, \textit{(ii)} different acquisition functions, and \textit{(iii)} the determination of the trust region center.
Afterwards (Section~\ref{sec:case_study}), selected methods are applied in a case study to maximize the traffic frequency on a realistic railway junction with eight possible routes.
All the following baselines are implemented in BoTorch~\cite{balandatBoTorchFrameworkEfficient2020} and to evaluate the CTMCs for all constraints we use \cite{Hensel.2022}.

\subsection{Ablation and Model Selection}
\label{sec:model_select}
In order to compare the different methods,
we consider a small exemplary network, consisting of three stations, connected via double-track lines (Fig. \ref{fig:example_junction}).
The routes from \textbf{A} to \textbf{C} ($r_1, r_3$) and from \textbf{A} to \textbf{D} ($r_2, r_4$) separate into two double-track railway lines in the railway junction \textbf{B}.

Formulating the queuing process of this junction necessitates the knowledge of minimum headway times in order to estimate the service rates for every route.
Usually, infrastructure managers would obtain them by applying some microscopic calculation tool based on infrastructure layout and driving dynamics (see \eg \cite{hansenRailwayTimetablingOperations2014}).
For this artificial example, however, we assume minimum headway times $h_{o, o'}$ as in Table \ref{tab:min_hw_ms}, with first trains $o$ in rows and following trains $o'$ in columns for the infrastructure in Fig. \ref{fig:example_junction}. Trains $(r_i,u) \in \Occs$ are indicated as `$r_i$--$u$', using the abbreviations `fr' for freight, `ld' for long-distance and `lo' for local trains to represent the different rolling stock units $u \in \mathcal{U}$.

\begin{table}[htbp]
\begin{center}
\caption{Minimum Headway Times in Minutes for the Junction Infrastructure in Fig.~\ref{fig:example_junction}}%
\label{tab:min_hw_ms}
\small
\begin{tabular}{lll l l l l l l l l l l l}
\toprule
 & &  & $r_1$ &  &  & $r_2$ &  &  & $r_3$ & & & $r_4$ & \\
 & & -fr & -ld & -lo & -fr & -ld & -lo & -fr & -ld & -lo & -fr & -ld & -lo \\
\midrule\midrule
&-fr & \cellcolor{lightgray}5.0 & \cellcolor{lightgray}5.0 & \cellcolor{lightgray}5.0 & 0.0 & 0.0 & 0.0 & \cellcolor{lightgray}5.0 & \cellcolor{lightgray}5.0 & \cellcolor{lightgray}5.0 & 0.0 & 0.0 & 0.0 \\
$r_1$&-ld & \cellcolor{lightgray}2.0 & \cellcolor{lightgray}2.0 & \cellcolor{lightgray}2.0 & 0.0 & 0.0 & 0.0 & \cellcolor{lightgray}2.0 & \cellcolor{lightgray}2.0 & \cellcolor{lightgray}2.0 & 0.0 & 0.0 & 0.0 \\
&-lo & \cellcolor{lightgray}3.0 & \cellcolor{lightgray}4.0 & \cellcolor{lightgray}3.0 & 0.0 & 0.0 & 0.0 & \cellcolor{lightgray}3.0 & \cellcolor{lightgray}4.0 & \cellcolor{lightgray}3.0 & 0.0 & 0.0 & 0.0 \\
\midrule
&-fr & 0.0 & 0.0 & 0.0 & \cellcolor{lightgray}5.0 & \cellcolor{lightgray}5.0 & \cellcolor{lightgray}5.0 & 1.7 & 1.7 & 1.7 & \cellcolor{lightgray}5.0 & \cellcolor{lightgray}5.0 & \cellcolor{lightgray}5.0 \\
$r_2$&-ld & 0.0 & 0.0 & 0.0 & \cellcolor{lightgray}2.0 & \cellcolor{lightgray}2.0 & \cellcolor{lightgray}2.0 & 1.0 & 1.0 & 1.0 & \cellcolor{lightgray}2.0 & \cellcolor{lightgray}2.0 & \cellcolor{lightgray}2.0 \\
&-lo & 0.0 & 0.0 & 0.0 & \cellcolor{lightgray}3.0 & \cellcolor{lightgray}4.0 & \cellcolor{lightgray}3.0 & 1.0 & 1.3 & 1.0 & \cellcolor{lightgray}3.0 & \cellcolor{lightgray}4.0 & \cellcolor{lightgray}3.0 \\
\midrule
&-fr & \cellcolor{lightgray}5.0 & \cellcolor{lightgray}5.0 & \cellcolor{lightgray}5.0 & 1.7 & 1.7 & 1.7 & \cellcolor{lightgray}5.0 & \cellcolor{lightgray}5.0 & \cellcolor{lightgray}5.0 & 0.0 & 0.0 & 0.0 \\
$r_4$&-ld & \cellcolor{lightgray}2.0 & \cellcolor{lightgray}2.0 & \cellcolor{lightgray}2.0 & 1.0 & 1.0 & 1.0 & \cellcolor{lightgray}2.0 & \cellcolor{lightgray}2.0 & \cellcolor{lightgray}2.0 & 0.0 & 0.0 & 0.0 \\
&-lo & \cellcolor{lightgray}3.0 & \cellcolor{lightgray}4.0 & \cellcolor{lightgray}3.0 & 1.0 & 1.3 & 1.0 & \cellcolor{lightgray}3.0 & \cellcolor{lightgray}4.0 & \cellcolor{lightgray}3.0 & 0.0 & 0.0 & 0.0 \\
\midrule
&-fr & 0.0 & 0.0 & 0.0 & \cellcolor{lightgray}5.0 & \cellcolor{lightgray}5.0 & \cellcolor{lightgray}5.0 & 0.0 & 0.0 & 0.0 & \cellcolor{lightgray}5.0 & \cellcolor{lightgray}5.0 & \cellcolor{lightgray}5.0 \\
$r_4$&-ld & 0.0 & 0.0 & 0.0 & \cellcolor{lightgray}2.0 & \cellcolor{lightgray}2.0 & \cellcolor{lightgray}2.0 & 0.0 & 0.0 & 0.0 & \cellcolor{lightgray}2.0 & \cellcolor{lightgray}2.0 & \cellcolor{lightgray}2.0 \\
&-lo & 0.0 & 0.0 & 0.0 & \cellcolor{lightgray}3.0 & \cellcolor{lightgray}4.0 & \cellcolor{lightgray}3.0 & 0.0 & 0.0 & 0.0 & \cellcolor{lightgray}3.0 & \cellcolor{lightgray}4.0 & \cellcolor{lightgray}3.0 \\
\bottomrule
\end{tabular}
\end{center}
\end{table}

In detail, we compare eight variants that differ across three key components: \textit{(i)} the acquisition function used, \textit{(ii)} the underlying mean function of the GP for capacity constraints, and \textit{(iii)} the trust region definition approach---either the conventional TuRBO framework (\texttt{TR}) or our proposed adaptation with re-centering to the probable optimum (\texttt{CPO}).
Table~\ref{tab:methods_defs} gives an overview regarding the different methods, their characteristics and their respective names.

\begin{table}[h]
\centering
\small
\caption{Proposed Variants and Their Defining Properties}%
\label{tab:methods_defs}
\begin{tabular}{llll}
\toprule
 & Acquisition & GP Mean & Trust Region \\ 
 Name    & Function    & Function  & Method \\
\midrule
\midrule
\texttt{EI-C-TR}   & EI & Const.     & Conv. TuRBO \\ 
\texttt{EI-C-CPO}      & EI & Const.     & Centered prob. Opti. \\ 
\texttt{EI-Exp-TR} & EI & Exp       & Conv. TuRBO \\ 
\texttt{EI-Exp-CPO}    & EI  & Exp       & Centered prob. Opti. \\
\midrule
\texttt{TS-C-TR}   & TS  & Const.     & Conv. TuRBO \\ 
\texttt{TS-C-CPO}      & TS  & Const.     & Centered prob. Opti. \\ 
\texttt{TS-Exp-TR}   & TS  & Exp       & Conv. TuRBO\\ 
\texttt{TS-Exp-CPO}    & TS  & Exp       & Centered prob. Opti.\\
\bottomrule
\end{tabular}
\vspace{-0.5em} 
\end{table}

In this example, we utilize the objective function from \eqref{eq:obj_general} for a given target parameter 
\begin{equation}
    \tilde{p}_u =\begin{cases} 
            0.5 &, ~u = \text{local train} \\
            0.3 &, ~u = \text{freight train}\\
            0.2 &, ~u = \text{long distance train}
\end{cases}
\end{equation}
of the traffic type distribution across the junction.

We conducted 20 experiments for all eight variants, using a weight factor of $w_U = 5$.
All computations were done on eight cores of an Intel Xeon Platinum 8468 Sapphire Processor (2.1 GHz), utilizing 19.84 GB of working memory. 
The termination criteria included a time limit of 60 min or a maximum of 200 iterations.

\begin{figure*}[t]
    \centering
    \subfloat[][Objective per Iteration]{\includegraphics[width=0.8\textwidth]{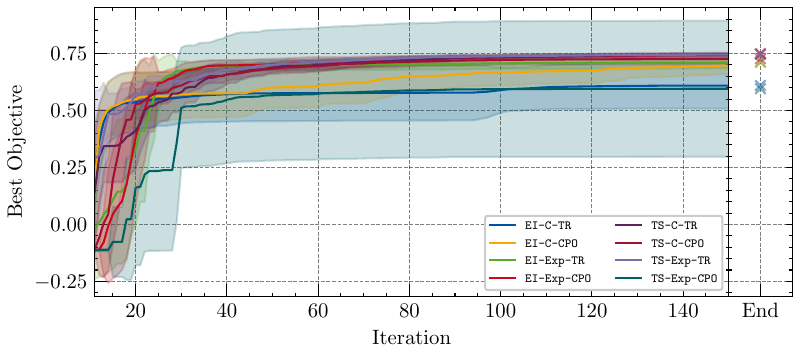}
        \label{fig:model_select_obj_distr_iter}}\\ \vspace{0.2cm}
    \subfloat[][Objective per Time]{
        \includegraphics[width=0.44\textwidth]{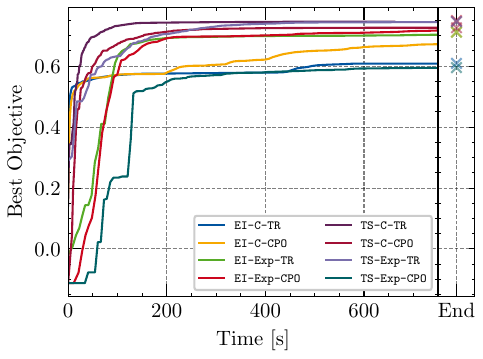}
        \label{fig:model_select_obj_distr_time}}
        \hfill
    \subfloat[][Objective at Termination]{
        \includegraphics[width=0.485\textwidth]{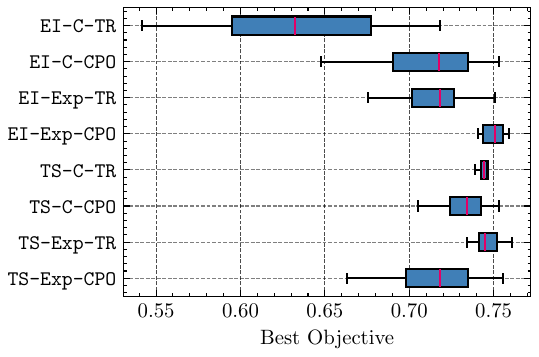}
        \label{fig:model_select_obj_distr_termin}}
    \caption{Objective distribution between all methods and experiments for $w_U = 5$.}
    \label{fig:model_select_obj_distr}
\end{figure*}

Fig.~\ref{fig:model_select_obj_distr} exemplarily illustrate the resulting objective function values for all variants. 
The results in Fig.~\ref{fig:model_select_obj_distr_iter} show the mean value and standard deviation $(\text{mean} \pm \sigma)$ of the best objective value at each iteration, averaged across 20 experiments per variant.
Initially, variants \texttt{EI-C-TR} and \texttt{EI-C-CPO} exhibit good performance during the first 20 iterations, achieving higher mean objective values. 
However, as the iterations progress, \texttt{EI-Exp-TR} and \texttt{EI-Exp-CPO} demonstrate improved efficacy and the best performance.
After approximately 50 iterations, methods \texttt{TS-Exp-TR}, \texttt{TS-C-CPO}, and \texttt{TS-C-TR} converge to similar mean objective values as those achieved by \texttt{EI-Exp-TR} and \texttt{EI-Exp-CPO}.
Upon termination of all methods (Fig.~\ref{fig:model_select_obj_distr_termin}), \texttt{EI-Exp-CPO}, \texttt{TS-Exp-TR}, and \texttt{TS-C-TR} achieve the highest mean objective values.
It is noteworthy that three TS methods---\texttt{TS-Exp-TR}, \texttt{TS-C-CPO}, and \texttt{TS-C-TR}---achieve their results more rapidly concerning actual computing time (Fig.~\ref{fig:model_select_obj_distr_time}) compared to the other methods.
This efficiency can be attributed to the reduced method overhead, resulting in faster iteration times. 
For this exemplary junction, the model-checking required in each iteration takes approximately $0.2~\text{s}$ to determine the capacity constraints (Section \ref{subsec:constr_formulation}) for the small railway junction analyzed. 
For more complex infrastructures, model-checking time may increase significantly, potentially diminishing the impact of method overhead on the overall computation time.
The following case study introduces a more complex railway junction infrastructure requiring more time for the formulation of capacity constraints in every iteration.

\subsection{Case Study on a Realistic Junction}
\label{sec:case_study}
In this Section, we study the performance of a railway junction, referred to the \textit{Triangle of Gagny}, located in the proximity of Paris, France.

\begin{figure}[t]
    \centering
    \includegraphics[width=0.75\linewidth]{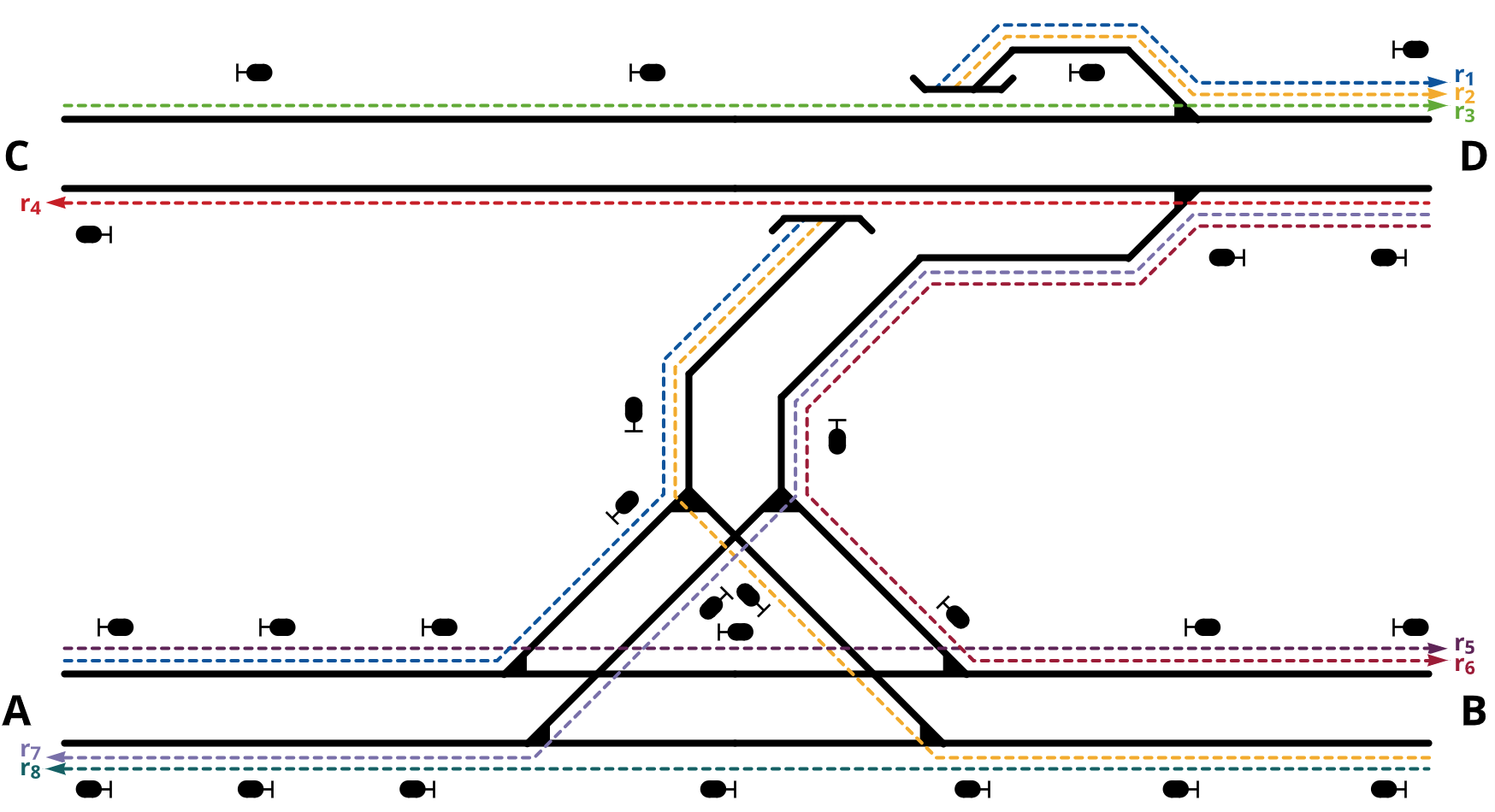}
    \caption{Infrastructure layout of the Triangle of Gagny. 
    Adapted from \cite{pellegriniOptimalTrainRouting2014,emundsEvaluatingRailwayJunction2024}}
    \label{fig:example_junction_gagny}
\end{figure}

This junction (Fig. \ref{fig:example_junction_gagny}) features eight routes between two double-track railway lines, \textbf{A}-\textbf{B} and \textbf{C}-\textbf{D}.
Similarly to the previous example, we consider straight routes as main routes and therefore have $\Routes_{\text{main}} = \{ r_3, r_4, r_5, r_8\}$.

This infrastructure has already been analyzed regarding rescheduling algorithms \cite{pellegriniOptimalTrainRouting2014} and static timetable capacity \cite{emundsEvaluatingRailwayJunction2024}.
Despite this, microscopic data is not openly available, we therefore assume the minimum headway times in Table~\ref{tab:min_hw_gagny}.

\begin{table}[htbp]
    \centering
    \small
    \caption{Minimum headway times in minutes for the Triangle of Gagny Junction (Fig.~\ref{fig:example_junction_gagny}).}
    \begin{tabular}{lllllllll}
        \toprule
         & $r_1$ & $r_2$ & $r_3$ & $r_4$ & $r_5$ & $r_6$ & $r_7$ & $r_8$ \\
        \midrule\midrule
        $r_1$ & 3.0 & 3.0 & 2.3 & 0.0 & 2.2 & 0.0 & 0.0 & 0.0 \\
        $r_2$ & 3.0 & 3.0 & 2.3 & 0.0 & 1.5 & 0.0 & 1.5 & 2.2 \\
        $r_3$ & 1.5 & 1.5 & 1.5 & 0.0 & 0.0 & 0.0 & 0.0 & 0.0 \\
        $r_4$ & 0.0 & 0.0 & 0.0 & 1.5 & 0.0 & 1.5 & 1.5 & 0.0 \\
        $r_5$ & 1.8 & 1.5 & 0.0 & 0.0 & 1.8 & 1.8 & 1.5 & 0.0 \\
        $r_6$ & 0.0 & 0.0 & 0.0 & 2.7 & 2.7 & 2.7 & 2.7 & 0.0 \\
        $r_7$ & 0.0 & 1.5 & 0.0 & 3.0 & 1.5 & 2.7 & 3.0 & 3.0 \\
        $r_8$ & 0.0 & 1.8 & 0.0 & 0.0 & 0.0 & 0.0 & 1.8 & 1.8 \\
        \bottomrule
    \end{tabular}

    \label{tab:min_hw_gagny}
\end{table}

Note that we restrict the case study to one type of trains, \ie local passenger trains.
Due to the route decomposition approach, this does not significantly reduce the required effort for the queue-length determination method from Section \ref{subsec:ql_estimations}.

Since this junction features eight routes, the CTMC for queueing-length estimations is more complex and requires substantially more computation time (\cf \cite{emundsEvaluatingRailwayJunction2024} for a more detailed scaling analysis). 
We therefore restrict the number of waiting positions to $B=3$ and the comparison to three good performing methods from Section \ref{sec:model_select} only: \texttt{EI-Exp-TR}, \texttt{EI-Exp-CPO} and \texttt{TS-C-TR}.
Furthermore, a time limit of $60$ minutes has been set, in order to test the real-world applicability of the methods.
For each of the three methods, 20 experiments have been computed for every selected weighting parameter $w_R$ on six cores of an Intel Xeon Platinum 8468 Sapphire Processor (2.1 GHz), utilizing 30.35 GB of working memory. 

Regarding the objective function, we assume that an infrastructure manager may be particularly interested in introducing a target parameter for the route distribution. 
We assume this target distribution to be
\begin{equation}
\label{eq:traffic_target}
    \tilde{p}_r =\begin{cases} 
            1/6 &, ~r\in \Routes_{\text{main}}\\
            1/12 &, ~ \text{otherwise}
\end{cases}
\end{equation}
which doubles the traffic on main routes.
The objective function has been formulated as
\begin{equation}
    \text{\eqrefOPT{eq:OPT-a}:} \quad \max_{\ovar \in \feasset}  \sum_{o \in \Occs} \ovar_o   - w_R \sum_{r \in \Routes}  \cdot \left( p_r - \tilde{p}_r\right)^2~.
\end{equation}
for different values of the weighting parameter $w_R \in \{0,1,2,5,10,20,50,100\}$.

Within the one hour time limit, the methods were able to perform between $30$ and $70$ iterations
On average, the algorithms were able to perform approximately $45$ iterations, while the first $10$ iterations were used as an initial training budget and their respective query points $\ovar$ were randomly selected from a pseudo-random Sobol sequence. 

\begin{figure*}[t]
    \centering
    \subfloat[][Objective per Iteration]{\includegraphics[width=0.455\textwidth]{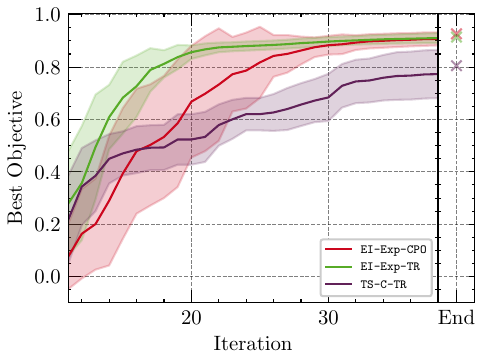}
        \label{fig:gagny_obj_iter}}
    \subfloat[][Total traffic volume]{
        \includegraphics[width=0.45\textwidth]{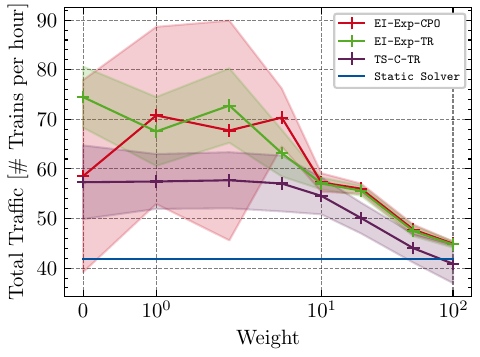}
        \label{fig:gagny_traffic_sum_weight}}
        \hfill
    \caption{
    Results for the modeled junction from Fig.~\ref{fig:example_junction_gagny}. Objective distribution per iteration for $w_R = 5$ 
    (\ref{fig:gagny_obj_iter}) and total traffic volume over different weights (\ref{fig:gagny_traffic_sum_weight}). The total traffic volume with our variants is compared to a solution from a static solver.}
    \label{fig:gagny_objs}
\end{figure*}

A comparison of the performance over iterations of the three selected variants is shown in Fig.~\ref{fig:gagny_obj_iter} for a weighting factor of $w_R = 5$.
In comparison with the results from Section~\ref{sec:model_select}, the variant \texttt{EI-Exp-TR} utilizing the base TuRBO framework outperforms the other two methods clearly in early stages.
However, \texttt{EI-Exp-CPO} is able to retrieve similar results at the end of the computation period---only \texttt{TS-C-TR} shows significant worse performance demonstrating that the additional prior knowledge in the GP model also here helps to significantly improve performance.
Note that these results are exemplary for all other experiments with different $w_R$ values.

In order to compare the total traffic sum across methods, Fig.~\ref{fig:gagny_traffic_sum_weight} displays the median and confidence intervals of the traffic volume for all selected weights.
Furthermore, the timetable capacity of $n_{\text{max}} \approx 41.92$ trains per hour for a static traffic distribution \eqref{eq:traffic_target}, calculated by solving the one-dimensional problem in \cite{emundsEvaluatingRailwayJunction2024}, determining only the total number of trains, has been denoted.
%
%
While all three methods exhibit large deviations for $w_R < 5$, relatively stable results have been obtained for weights $w_R > 5$.
In particular, variant \texttt{TS-C-TR}, using TS instead of EI, shows relatively constant standard deviations of 5 to 7 trains per hour.
In comparison, \texttt{EI-Exp-CPO} exhibits standard deviations of up to 20 trains per hour for small $w_R$, but under one train per per hour for $w_R = 100$.
Note that the lowest standard deviations are obtained with \texttt{EI-Exp-TR}, between $7.2$ and $0.6$ trains per hour.

Comparing the optimal traffic volume (and objective), \texttt{EI-Exp-TR} and \texttt{EI-Exp-\allowbreak{}CPO} show similar results for $w_R \geq 10$, finding solutions with 2 to 8 trains per hour more than \texttt{TS-C-TR}.
Their medians are, depending on $w_R$, between 30 and $2.5$ trains per hour higher, than the timetable capacity of $n_{\text{max}} \approx 41.92$ trains per hour for a static traffic distribution.

Higher weight factors $w_R$ enforce smaller deviations from the in \eqref{eq:traffic_target} specified route distribution $\tilde{p} = (\tilde{p}_r)_{r \in \Routes}$.
In the following, we analyze the \textit{penalty} or distance $\text{dist}(p, \tilde{p})$ between the route distribution $p=(p_r)_{r \in \Routes}$ for a fixed solution $\ovar$ to the specified $\tilde{p}$.
We utilize the Euclidean metric $\text{dist}(p, \tilde{p}) = \sqrt{\sum_{r\in \Routes}(p_r - \tilde{p}_r)^2} $ as the distance here.

\begin{wrapfigure}{r}{0.5\textwidth}
    \centering
    \vspace{-1em} 
    \includegraphics[width=0.48\textwidth]{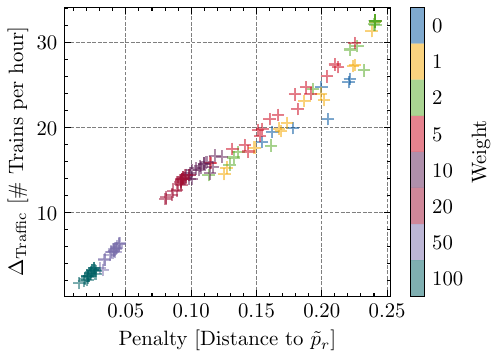}
    \caption{Traffic volume difference and penalty of all obtained solutions for \texttt{EI-Exp-TR}.}
    \label{fig:gagny_traffic_diff}
    \vspace{-1em} 
\end{wrapfigure}

Based on all obtained solutions with method \texttt{EI-Exp-TR}, Fig.~\ref{fig:gagny_traffic_diff} compares this penalty to the traffic volume difference $\Delta_{\text{Traffic}} = n(\ovar^{\ast}) - n_{\text{max}}$ between the  traffic volume $n(\ovar^{\ast})$ for the obtained solution $\ovar^{\ast}$ and the static timetable capacity $n_{\text{max}}$.


In accordance with the previously discussed results, deviations between solutions are particularly high for lower weights $w_R < 10$, and more distinct clusters can be recognized for higher weights.

However, the results indicate a significant correlation between $\Delta_{\text{Traffic}}$ and the penalty of the obtained solutions for \texttt{EI-Exp-TR}.
For solutions close to the given distribution with small penalties, less traffic volume can be gained than for solutions with higher distances to $\tilde{p}$.
Furthermore, instances with small weights $w_R$ exhibit far more variation in their balance between adherence to the given route distribution and achieved traffic volume.

These observed effects also follows intuition on the influence of the weight.
As the weight increases, the optimization problem \OPT becomes more regularized, smoothing the landscape of the objective. 
Conversely, smaller weights yield more flexibility at the cost of multiple local minima.

Therefore, infrastructure managers, applying the introduced methods for long-term planning problems, should carefully select the highest  weight factor suitable for their exact needs, in order to take advantage of the enhanced convergence and stability of solutions for increased penalty weights.

\section{Discussion and Conclusion}\label{sec:discussion}
In this work, a novel method to determine the performance capability of a railway junction under consideration of dynamic traffic distributions has been introduced.
Extensions to an existing queueing-based method have been developed (Section \ref{sec:qs_all}) by implementing uncertainties within capacity constraints of a traffic rate assignment problem.
The introduced method utilizes BO in order to efficiently solve the presented penalized traffic rate assignment problem \OPT from Section~\ref{sec:problem}.
For this, in Section~\ref{sec:bo}, we tailored standard BO algorithms to the setting of known objective under black-box constraints and included domain knowledge in the GP surrogates to increase sample efficiency.
On an exemplary infrastructure of a small railway junction, we performed an ablation of our approaches in Section~\ref{sec:model_select} of which we selected the most promising candidates for further investigation.
These selected variants have been applied to a complex junction infrastructure adapted from a real-world example in Section~\ref{sec:case_study}, highlighting the method's ability to assess infrastructure capacity and further quantifying the impact of allowing deviations from traditionally static traffic distributions.

Within the case studies, the real-world applicability of the proposed methods becomes evident, especially when considering scenarios with a high emphasis on small deviations from enforced distributions.
However, in the case of small regularization weights, results show high variability for methods utilizing EI on complex infrastructure examples.
This indicates that small weights can lead to optimization problems with multiple local optima, which pose a challenge to the introduced methods, iteratively improving their solution quality in the neighborhood of the last best solution.
Even for high regularization weights, obtained results are not guaranteed to form a global optimum to the posed traffic rate assignment problem.
Still, all variants outperformed the static solution, indicating a clear potential of adapting the proposed methods in practice.
Future work could leverage parametric model-checking alongside classical non-linear solvers with multiple random initializations to efficiently locate promising local optima when solving \OPT.
Further extensions may include the adaption of capacity constraints to reflect other capacity metrics or the performance analysis of entire railway networks, utilizing element-wise capacity constraints.



To conclude, the BO variants developed in this work efficiently obtain feasible, locally optimal solutions for the capacity-constrained traffic rate assignment problem under dynamic traffic distributions.
Our framework of optimizing known objectives subject to unknown constraints represents a novel contribution to the field of BO with potential applications across multiple domains beyond railway applications where similar problem structures arise.
For our specific problem, it enables infrastructure managers to specify acceptable variations in traffic composition when estimating the timetable-independent capacity of railway junctions, making it particularly well-suited for long-term infrastructure planning scenarios where future traffic distributions across train types or lines remain uncertain.



\section*{Acknowledgements}
This work is funded by the Deutsche Forschungsgemeinschaft (DFG, German Research Foundation) – 2236/2. Computational Experiments were performed with computing resources granted by RWTH Aachen University under project rwth1635.

\section*{CRediT authorship contribution statement}
\textbf{Tamme Emunds:} Conceptualization, Methodology, Software, Formal Analysis, Writing –  original draft. \textbf{Paul Brunzema:} Conceptualization, Methodology, Software, Formal Analysis, Writing –  original draft. \textbf{Sebastian Trimpe:} Funding Acquisition, Supervision, Writing - review \& editing. \textbf{Nils Nießen:} Funding Acquisition, Supervision, Writing - review \& editing.

\bibliographystyle{elsarticle-num}
\bibliography{references}

\newpage
\appendix
\section{Notation}
\label{app:vars}


\begin{longtable}{@{}cll@{}}
\caption{Overview of Variables and Notation}
\label{tab:variables}\\
\toprule
\textbf{Symbol} & \textbf{Type} & \textbf{Description} \\
\midrule
\endfirsthead
\caption{Overview of Variables and Notation}\\
\toprule
\textbf{Symbol} & \textbf{Type} & \textbf{Description} \\
\midrule
\endhead
\multicolumn{3}{l}{\textit{Infrastructure and Network}} \\
$J$ & Set & Railway junction $J = (\mathcal{R}, C)$ \\
$\mathcal{R}$ & Set & Set of $k$ routes in junction \\
$C$ & Matrix & Conflict matrix $C \in \{0,1\}^{k \times k}$ \\
$\mathcal{U}$ & Set & Set of rolling stock units \\
$\mathcal{O}$ & Set & Set of occupation requests $o = (r, u) \in \mathcal{R} \times \mathcal{U}$ \\
\midrule
\multicolumn{3}{l}{\textit{Traffic and Capacity}} \\
$\optvar_o$ & Scalar & Arrival rate for request $o$ \\
$\lambda_r$ & Scalar & arrival rate for route $r$, $\lambda_r = \sum_{(r,u) \in \Occs} \ovar_{(r,u)}$ \\
$\optvar$ & Vector & Assignment vector $(\optvar_o)_{o \in \mathcal{O}} \in \mathbb{R}^d$ \\
$f_o$ & Scalar & Frequency of trains per request $o$ \\
$t_U$ & Scalar & Fixed time horizon \\
$d$ & Scalar & Dimension of assignment vector, $d = |\mathcal{O}|$ \\
\midrule
\multicolumn{3}{l}{\textit{Service Times and Operations}} \\
$h_{o,o'}$ & Function & Minimum headway time between requests $o$ and $o'$ \\
$b_o$ & Scalar & Average occupation time for request $o$ \\
$b_r$ & Scalar & Average occupation time for route $r$ \\
$\mu_r$ & Scalar & Service rate for route $r$, $\mu_r = 1/b_r$ \\
$\rho_r$ & Scalar & Occupation ratio for route $r$, $\rho_r = \optvar_r/\mu_r$ \\
$v_A, v_S$ & Scalar & Coefficients of variation of the arrival and service process \\
\midrule
\multicolumn{3}{l}{\textit{Optimization Problem}} \\
$f(\optvar)$ & Function & Objective function \\
$\Lambda$ & Set & Feasible set \\
$\optvar^*$ & Vector & Optimal solution \\
$ub_o$ & Function & Upper bound function for request $o$ \\
$p_u(\optvar)$ & Function & Traffic distribution for train type $u$ \\
$p_r(\optvar)$ & Function & Traffic distribution for route $r$ \\
$\tilde{p}_u$, $\tilde{p}_r$ & Scalar & Target distributions for types/routes \\
$w_U$, $w_R$ & Scalar & Weighting factors for distribution penalties \\
\midrule
\multicolumn{3}{l}{\textit{Constraint Formulation}} \\
$L_r$ & Scalar & Expected queue length for route $r$ \\
$L_{\text{limit},r}$ & Scalar & Threshold for expected queue length on route $r$ \\
$c_r(\optvar)$ & Function & Capacity constraint function for route $r$ \\
$p_{pt,r}$ & Scalar & Proportion of passenger traffic on route $r$ \\
\midrule
\multicolumn{3}{l}{\textit{Bayesian Optimization}} \\
$m(\optvar)$ & Function & GP mean function \\
$k(\optvar, \optvar')$ & Function & GP kernel function \\
$\alpha(\optvar)$ & Function & Acquisition function \\
$\mathcal{D}_t$ & Set & Dataset at iteration $t$ \\
$\mathcal{TR}_t$ & Set & Trust region at iteration $t$ \\
$L_t$ & Scalar & Length of the TuRBO hyperbox at iteration $t$\\
$\mathcal{F}$ &Set& Feasible points in candidates set \\
$\mathcal{M}$ & Set & Candidates for selection \\
\midrule
\multicolumn{3}{l}{\textit{GP Model Parameters}} \\
$\beta$, $w_i$, $\gamma$ & Scalar & Exponential mean function parameters \\
$\sigma_f^2$ & Scalar & GP output scale \\
$\nu$ & Scalar & Parameter for the Mat\'ern kernel \\
$\ell_i$ & Scalar & Length scale for dimension $i$ \\
$\sigma_n^2$ & Scalar & Noise variance \\
$\Theta_{r}$ & Set & Set of learnable hyperparameters \\
\midrule
\multicolumn{3}{l}{\textit{General notation}} \\
$\mathcal{I}_T $, $\mathcal{I}_d $  & Set & Set of first $T$/$d$ positive integers \\
$\Delta$ & Function & Euclidean distance \\
$\sigma$ & Scalar & Standard Deviation \\
\bottomrule
\end{longtable}
%

\end{document}